\input psfig.sty
\documentclass{aa}
\begin{document}

   \thesaurus{08.16.7}  
   
   \title{An Atmospheric Model for the Ion Cyclotron Line of Geminga. 
}

\author{A. Jacchia\inst{1}, F. De Luca\inst{1,2}, E. Lazzaro\inst{1}, 
P. A. Caraveo\inst{3,4}, R.P. Mignani\inst{5}  and
G.F. Bignami\inst{6,7}} 

   \offprints{A. Jacchia }

\institute{Istituto di Fisica del Plasma ``P. Caldirola'', CNR, Via Cozzi 
53, 20125-Milan, Italy \\
email:jacchia@ifp.mi.cnr.it 
\and 
Universit\'a degli studi di Milano, Dipartimento di Fisica, Via
Celoria 16, 20133-Milan, Italy
\and         
Istituto di Fisica Cosmica ``G. Occhialini'', CNR, Via Bassini 15, 20133-Milan, Italy 
\and 
Istituto Astronomico, Via Lancisi 29, 00161-Rome, Italy 
\and 
STECF-ESO, Karl Schwarzschild Str.2,  D85740-Garching, Germany 
\and
Agenzia Spaziale Italiana, Via di Villa Patrizi 13, 00161-Rome, Italy 
\and
Universit\'a di Pavia, Via Bassi 6,  27100-Pavia, Italy 
         }

 \date{Received ; accepted  }

\authorrunning{Jacchia et al.}
\titlerunning{An Atmospheric Model for the Ion Cyclotron Line of Geminga.}

\maketitle

\begin{abstract}
 
Recent    optical/UV data  of  Geminga  (Mignani   et al.  1998)  have
strengthen the  significance of the emission  feature present over the
thermal continuum   best fitting the   EUV/soft  X-ray data.   Here we
present a phenomenological model interpreting the  data in terms of an
ion cyclotron feature, arising from  a thin outer  layer of hot plasma
covering  Geminga's polar caps.  The width  of the feature favours for
Geminga the strongly oblique rotator configuration, as already deduced
from   the  soft  X-ray  data  (Halpern   \&  Ruderman   1993).   Our
interpretation   implies a  magnetic field  of   $3 - 5~  10^{11}~ G$,
consistent with the value deduced from the dynamical parameters of the
pulsar.    This  would represent  the  first   case  of an independent
measurement of   the surface  magnetic field  of   an isolated neutron
star.  \\ The expected  time modulation  of  Geminga's feature is also
discussed. 

   \end{abstract}

%

\section{Introduction}

The panorama  of optical observations of  Isolated Neutron Stars (INS)
is limited by  the faintness of the vast  majority of  them.  Only one
object, the Crab pulsar,  has good, medium resolution  optical (Nasuti
et al. 1996)  and near-UV spectral data (Gull  et al. 1998), while for
PSR0540-69 the synchrotron continuum has been measured by HST (Hill et
al.  1997).  For a  few  more cases  acceptable multicolour photometry
exists, while the rest of the data base (a grand total of less than 10
objects currently) consists  of one/two - wavelengths detections  (see
Caraveo 1998 for a  summary of the  observational panorama).  \\ Apart
from the very young objects,  characterized by flat,  synchrotron-like
spectra arising   from   energetic electron  interactions    in  their
magnetosphere, of particular interest  are the middle-aged ones ($\sim
10^{5}~ yrs$   old).  The non-thermal,  magnetospheric emission should
have faded enough (in the X-ray waveband at least) to render visible
the  thermal emission  from  the  hot INS  surface.   Standard cooling
calculations    predict  a    surface   temperature   in   the   range
$10^{5}-10^{6}~^{\circ}K$, in  excellent  agreement  with recent X-ray
observations of  INS with thermal  spectra (e.g.  Becker \& Tr\"umper
1997).  It is easy to predict the IR-optical-UV fluxes generated along
the $\sim E^{2}$ Rayleigh-Jeans slope of the Planck curve best fitting
the X-ray data, and   to  compare predictions to  observations,  where
available.  \\ In what follows, we shall concentrate on Geminga, which
is certainly the most studied object of its class (Bignami \& Caraveo
1996 and refs.   therein), and possibly   of all INSs (except for  the
Crab).   Its IR-optical-UV data (Bignami et  al. 1996;  Mignani et al.
1998) show the presence  of a  well defined  emission feature. Such  a
feature is superimposed  on  the thermal continuum expected  from  the
extrapolation  of the  black-body  X-ray  emission detected by   ROSAT
(e.g. Halpern \&  Ruderman 1993).  The  presence  of a  clear maximum
centered on V has been recently  questioned by Martin  et al. (1998) on
the basis of a spectrum which is at the limit of the capability of the
Keck telescope.  The spectrum of Geminga,  detected at a level of just
0.5\% of   the  dark sky,  seems fairly  flat   and, although  broadly
consistent with earlier measurements,  it is definitely above the flux
measured   in the B-band by     the HST/FOC (Mignani   et al.   1998).
Therefore,  in  view  of the    faintness  of the   target,  we  shall
concentrate on the  photometric measurements, which have been repeated
using different instrumental set-ups and appear more reliable than the
available spectral data.  \\  Recently, pulsations in the B-band  have
been tentatively detected by Shearer et al.  (1998). Also in this case,
the faintness of  the source limits quite severely  the S/N  ratio and
thus   the statistical significance of the   result.  Indeed, a pulsed
signal at just the 3.5 $\sigma$ level was found during only one of the
three nights devoted to the project.  If confirmed, these measurements
would   have deep implications on  the  mechanisms responsible for the
optical  emission of  Geminga.  However,   in  view of the  rather low
statistical significance  of these   results,  we shall  stand by  the
interpretation   of Mignani    et   al.    (1998)  and  propose     a
phenomenological  model  interpreting the   feature as  an atmospheric
cyclotron line emission from Geminga's polar caps.


\section{The Spectral Distribution}

\begin{figure}
\centerline{\hbox{\psfig{figure=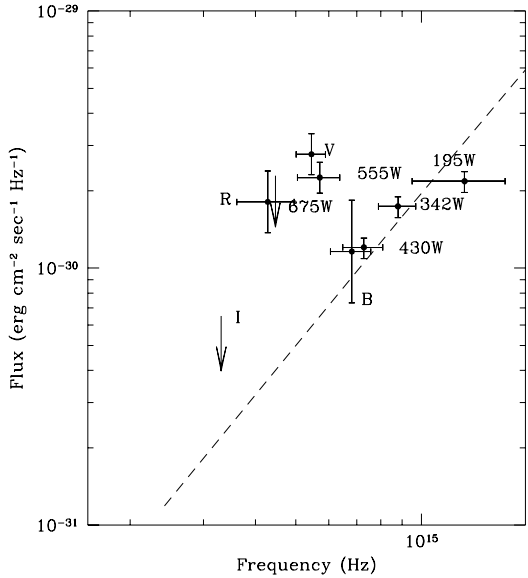,height=8cm,clip=}}}
\caption{The plot summarizes the complete multiband HST and
ground-based photometry  of  Geminga. Three digits  identify WFPC2/FOC
imaging filters.  The dashed line  represents the extrapolation in the
optical of the blackbody  fit to the ROSAT  data (Halpern \& Ruderman
1993),   which,  for     the     best   fitting   temperature       of
$T=5.77~10^{5}~^{\circ}K$ and the measured distance $d \simeq 157~ pc$
(Caraveo et al. 1996), yield an emitting radius $R=10~km$.} 
\end{figure}

Fig.1 shows the complete I-to-near UV colours  of Geminga.  These were
obtained   both from the  ground  and  from  HST  during  ten years of
continuing   observational effort   (from   Mignani et    al.   1998).
Comparing the repeteadly confirmed V-band  magnitude with the relative
fluxes of the three FOC  points (430W, 342W and 195W)  in the B/UV and
the  upper limit in  I, it  is easy to   recognize an emission feature
superimposed  to a $\sim  E^{2}$ RJ-like continuum.  It is thus
obvious to   directly compare  the measured   optical   fluxes to  the
extrapolation of the soft X-ray   blackbody spectrum.  However,   this
apparently  trivial step is   easier  said than done, since  different
X-ray  observations have yielded  for  Geminga slightly different best
fitting temperatures,  with significantly different bolometric fluxes.
For example, two independent  ROSAT observations yielded  best fitting
temperatures of   $5.77~10^{5}~^{\circ}K$ (Halpern \&  Ruderman 1993)
and  $4.5 ~10^{5}~^{\circ}K$ (Halpern   \& Wang  1997), respectively,
implying, for the same  Geminga distance, a  factor of 3 difference in
the   emitting area.  As    a  reference, we    show  in  Fig. 1   the
Rayleigh-Jeans   extrapolation  of the  ROSAT X-ray  spectrum obtained
using the   best fitting temperature derived  by  Halpern  \& Ruderman
(1993) which, at the Geminga distance (Caraveo et al. 1996), yields an
emission radius of 10 km.  Thus, while the optical, RJ-like, continuum
is  largely consistent with   thermal emission from  the neutron  star
surface, the   feature around $\sim 6,000   \AA$, requires a different
interpretation.  \\ In  the following,  we propose a  phenomenological
model interpreting the   feature  seen  in Fig.1 as    an  atmospheric
cyclotron line emission from Geminga's polar caps.

\section{The Ion Cyclotron Emission Model.}

Comparisons between theory and  observed cooling  NS spectra have  been
performed, e.g.,  by Romani (1987)  and Meyer et al.  (1994). Basically,
the star surface is thought to be surrounded by a colder, partially ionized,
atmosphere. Such an atmosphere behaves as a broad-band, absorbing-emitting
medium. The  observed spectrum is due to   a rather complicate process,
involving transfer of radiating  energy  between regions of  different
depths, temperatures and chemical compositions. These models predict a
deviation from the blackbody  emission law marginally observed in  cooling
neutron star spectra.  Models that account for the effects of the star
magnetic  field $B$   foresee  absorption   lines  at   the  cyclotron
frequencies.  This  is easily  explained by   the presence of resonant
frequencies,  created by  the magnetic  field, at  which the radiation
emitted from   deeper regions is quite  efficiently   absorbed. On the
other hand,  a cyclotron emission  "line", superimposed to the blackbody
continuum, requires a thermal inversion of the stellar atmosphere.  We
will  see in the following  that a thin hot  plasma layer is optically
thick at the cyclotron  frequency, so that the cyclotron emission is far
more efficient than  Bremsstrahlung in the same spectral range.  
For reasonable value  of  the  plasma density,  therefore,
Bremsstrahlung can be neglected. Geminga's  emission "line" might then
be  explained as  an (ion)  cyclotron  emission from  a magnetoplasma,
consisting of  a mixture of Hydrogen  and Helium, surrounding the star
surface.  Any feature of the observed radiation spectrum of Geminga in
the region of the proton cyclotron frequency  is ultimately due to the
accelerated  motion of charged particles in  the magnetic and electric
fields of the rotating star.
We assume, however, the upper  part of the star
atmosphere to  be  a fully ionized  "classical"  plasma layer,  with a
Maxwellian ion distribution function. The plasma  layer is immersed in
the inhomogeneous star dipole magnetic field.  Accordingly, we develop
the theory of the emission under the condition of applicability of the
Kirchhoff's  law. To this  end  it is  necessary  to consider the full
dielectric response  of the  magnetized  plasma, and address  two  key
questions: (i) identification of an electromagnetic  (e.m.) wave which can
propagate  in vacuo toward the observer and assessment of its thermal
(black body) absorption/emission properties   in a single or  multiple
species plasma, (ii) assessment of  the emission line
broadening mechanism as a  consequence  of Doppler effect, collisions,
temperature and density inhomogeneity and averaging over the
magnetic field. 

\begin{figure}
{\bf (a)} \\
\centerline{\hbox{
\psfig{figure=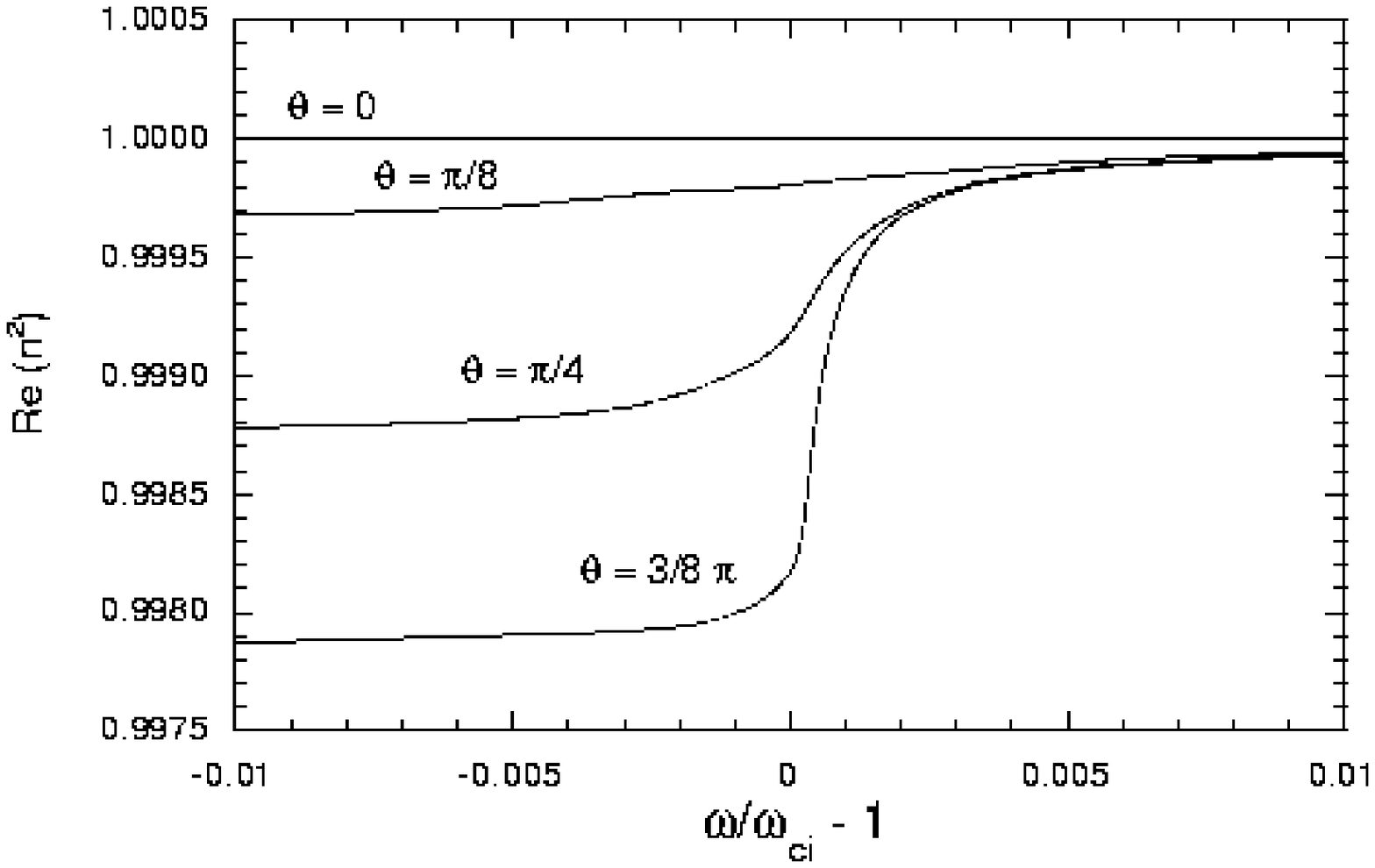,height=6cm,clip=}}}
{\bf (b)} \\
\centerline{\hbox{
\psfig{figure=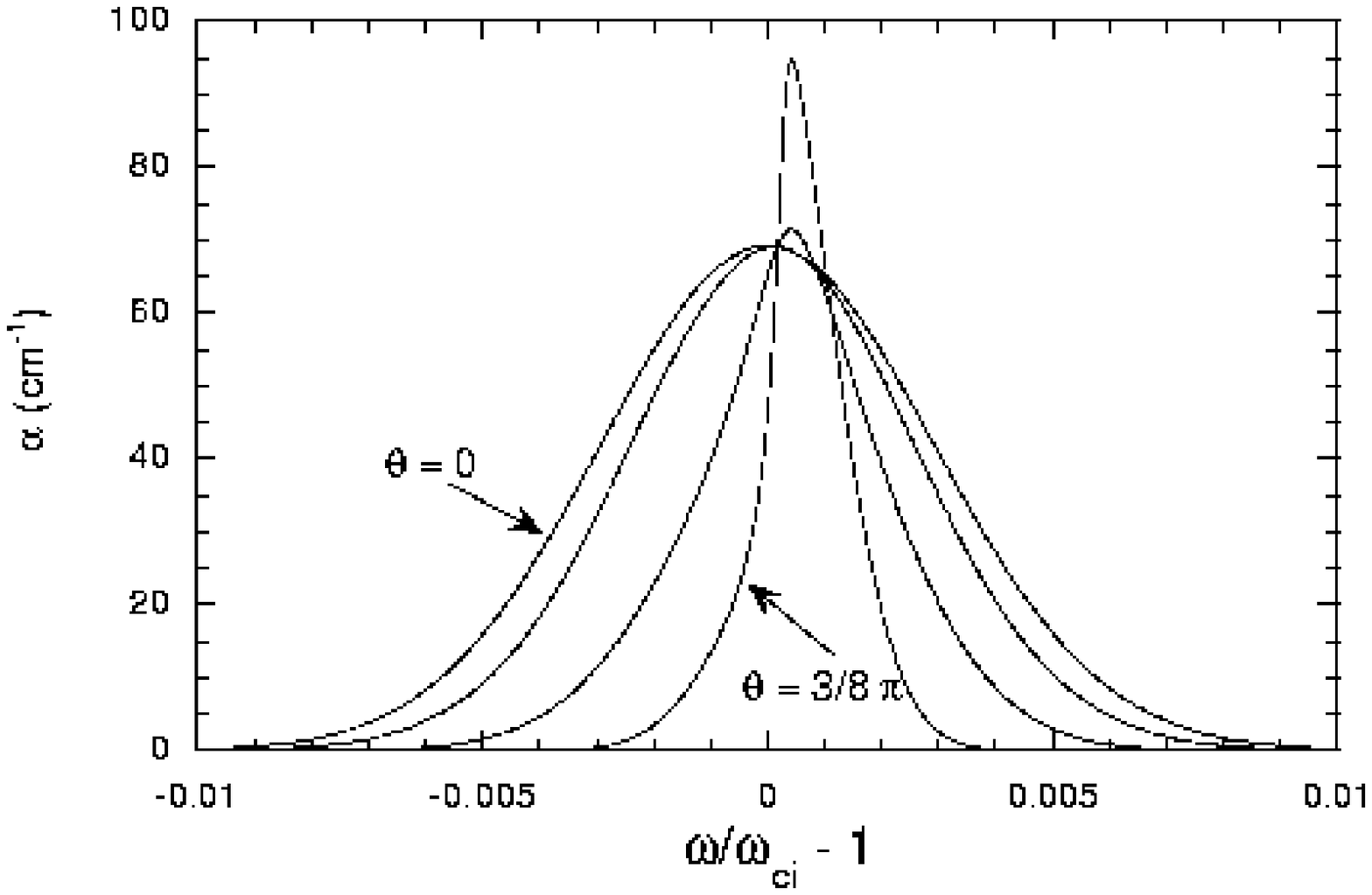,height=6cm,clip=}}}
\caption{
(a) Plot  of   the real part  of  the   square of refractive   index
$n^{2}(\omega,\theta)$  of   the fast wave   vs $\omega/\omega_{ci}-1$
(where $\omega_{ci}$ is  the ion  cyclotron frequency), for  different
values of the angle $\theta$ between the  direction of propagation and
the magnetic  field.  Propagation  toward  the observer, on   the "low
field side" $\omega/\omega_{ci} \ge 1$ is  guaranteed by the fact that
$n^{2}$ is positive.  (b)   Plot  of the absorption   coefficient
$\alpha(\omega,\theta)   [cm^{-1}]$    of    the    fast     wave   vs
$\omega/\omega_{ci} - 1$, for   different   values  of the   angle
$\theta$   between  the direction  of   propagation  and the  magnetic
field. Calculation is done for a Hydrogen Maxwellian plasma with $B = 
3.8~   10^{11}~ G$, $n =10^{19}[cm^{-3}]$, $T_{i}(\beta=0^{\circ})=
9~10^{7}~^{\circ}K$ (where $T_{i}$  is  the  ion temperature) and   it
shows that optical thickness is obtained for plasma layer of the order
of 1 cm at all propagation angles.} 
\end{figure}

\subsection{The Electromagnetic Wave and its Absorption/Emission Properties.}

If a medium  emits radiation at a  given frequency, it also absorbs it
at  the same frequency. The  quantity  $I(\omega)$, i.e. the  radiated
power per  unit area per unit solid  angle per unit angular frequency,
is given by the  solution of the radiative   transfer equation. For  a
single propagating e.m. mode in a medium of refractive index $n$ this is
written as: 

$$
n^{2} \frac{d}{ds} \Bigl( \frac{I(\omega)}{n^2} \Bigr) = 
j(\omega) - \alpha(\omega)I(\omega) \eqno (1)
$$

where  $j(\omega)$  and $\alpha(\omega)$  are  the  emissivity and the
absorption coefficient respectively and $s$ is the axis along the
radiation path. In  the present
problem  only one  normal  e.m. mode can   propagate in  vacuo  and be
observed.   The emissivity, $j(\omega)$,  is computed  from the plasma
dispersion relation and the Kirchhoff's law: 

$$
j(\omega) = \alpha(\omega)K(\omega,T) \eqno (2)
$$

where $K (\omega,T)$ is the Planck function at the temperature $T$,
where $T$ is here the plasma temperature.  
Near the  cyclotron frequency
$\omega_{ci}$ ($2\pi~  \nu_{ci}  = \omega_{ci} =  ZeB/A  c$; $Z$  is the
atomic charge  A  the atomic  mass), two  independent  e.m.  modes can
propagate in the magnetized plasma atmosphere.  They are identified by
an index of refraction given by the complex roots $n_{\pm}= n'+i~n''$ of
the complex dispersion relation written for a real frequency $\omega \simeq 
\omega_{ci}$ in the biquadratic form (Akhiezer et al. 1975): 

$$
An^{4}_{\pm} + Bn^{2}_{\pm} + C = 0 \eqno (3) 
$$

\begin{figure}
\centerline{\hbox{\psfig{figure=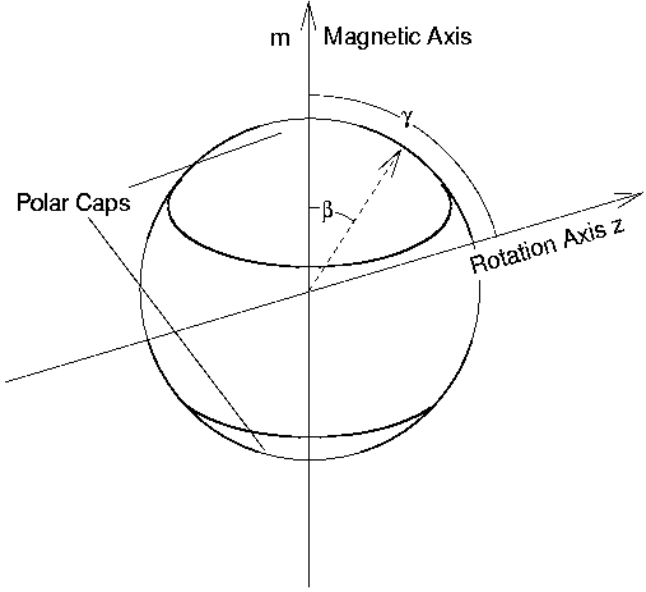,height=8cm,clip=}}}
\caption{Main geometrical parameters used in our model: $\gamma$
is the angle between the magnetic and the rotation axis. $\beta$ defines the
latitude of the polar caps.}
\end{figure}

\begin{figure}
\centerline{\hbox{\psfig{figure=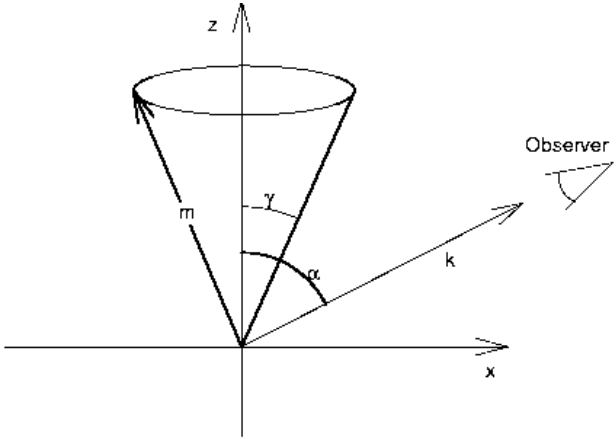,height=6cm,clip=}}}
\caption{Apart from $\gamma$ and $\beta$, our model call for a third angle,
$\alpha$, between the wave vector, pointing toward the observer, and the
rotation axis. The $\gamma$ angle between the magnetic and the rotation axis is
also shown.}
\end{figure}

The coefficients $A, B$  and $C$ are function  of the  full dielectric
tensor  and depend  on  the  plasma  frequency $\omega_{p}$,  the  ion
cyclotron frequency $\omega_{ci}$, the ion temperature $T_i$ and 
the angle $\theta$ between the propagation  wave vector  
$\vec k =  \frac{\omega}{c}  \vec n$ and the
local magnetic field    $\vec B$.  Of  these   two waves,   the slow  $n_+$
(ordinary)  wave diverges at   the  cyclotron resonance with  a finite
bandwidth (roughly given by the ratio  of the thermal velocity to the
phase  velocity)    of  anomalous  dispersion  around   the  cyclotron
frequency. 

The ordinary wave in the region where $\omega > \omega_{ci}$ 
has an evanescence gap, which even if
bridging by  thermal  and collisional effects  is considered, produces
substantial  attenuation of wave  propagating in tenuous plasma toward
the observer (Shafranov 1058, Akhiezer et.  al. 1975). The Geminga
plasma atmosphere is characterized by $\omega_{p}/\omega_{ci} < 1$. This prevents the use of the
customary ion cyclotron  approximation of the  dielectric tensor and a
numerical solution of the  dispersion relation  for both the  ordinary
and extraordinary  mode has been performed  . The numerical evaluation
of absorption and propagation  properties  of the mode propagating  in
vacuo  are shown in   Fig. 2.  as function of   the angle between  the
direction  of propagation and  the magnetic field .  The extraordinary
mode, which  can   propagate  in  vacuo  with frequency  $\omega   \ge
\omega_{ci}$, is the so called
magnetosonic, compressional Alfven or fast wave. 

It has   a  regular  index of
refraction and a very narrow band of frequency of anomalous dispersion
with a  correspondingly  narrow frequency range  of absorption  due to
thermal   effects   (Akhiezer  et  al,  l975).     The  fast  wave  is
electromagnetic in nature and in the cold plasma limit it is righthanded
polarized. It  becomes lefthanded (i.e.   in the direction  of the ion
motion), thus allowing absorption at  the fundamental harmonic, owing to
small thermal effects or to the presence in the plasma of small traces
of    isotopes   of the  minority  species.     The Geminga  plasma is
characterized by $\omega_{p}/\omega_{ci} < 1$. This prevents the use
of the customary ion cyclotron  approximation of the dielectric tensor
and   requires a numerical solution     of  the dispersion   relation.
Assuming an appropriate temperature  and  density, even  in a  pure $H^+$
plasma  the fast wave    reaches   a sufficiently  large    absorption
coefficient     $\alpha(\omega,\theta)       =  2  ~(\omega/c)~     n''
(\omega,\theta)$ .
The optical depth, $\tau$, can be estimated as

$$
\tau(\omega,\theta) = \int_0^L {\alpha(\omega,\theta) ds'} \simeq 
\alpha(\omega,\theta)  L \gg 1 ~ 
\eqno (4)
$$

where $L$ is the plasma layer thickness, of the order of few cm.    
The refraction index of the fast wave and its absorption
coefficient $\alpha$  are  given  in Fig. 2a,b  for different $\theta$   angles, a
plasma   density  of $10^{19}cm^{3}$    and  a plasma  temperature  of
$9~10^{7}~^{\circ}K$.  It   is worth noticing that,   owing  to  the weak
dependence of $\alpha$ on    $\theta$  (see  Fig.2b), a   layer   of
thickness   less  than 1  cm  is   sufficient to match   the blackbody
emission, independently of  the angle of  propagation with  respect to
the magnetic field at any point on the star  surface. We will use this
condition to greatly simplify the computation  of the intensity and of
the broadening of the emitted line. 

The intensity, $i(\omega,\omega_{c})$, radiated from any point of the star 
surface, is
then simply given
by

$$
i(\omega,\omega_{ci})= K(T,\omega) [1- e^{- \tau(\omega,\omega_{ci})}]
\eqno (5)
$$

A density limit of  about $10^{19}$ particles $cm^{-3}$  represents an upper
limit to the plasma density. Above such limit, bremsstrahlung, with its
absorption coefficient $\alpha \simeq 10^{-2} (cm^{-1})$ (Bekefi
1966) in the range of temperature
frequency of interest here, would
introduce severe spectral distortions, not observed in the  data.

\begin{figure}
\centerline{\hbox{\psfig{figure=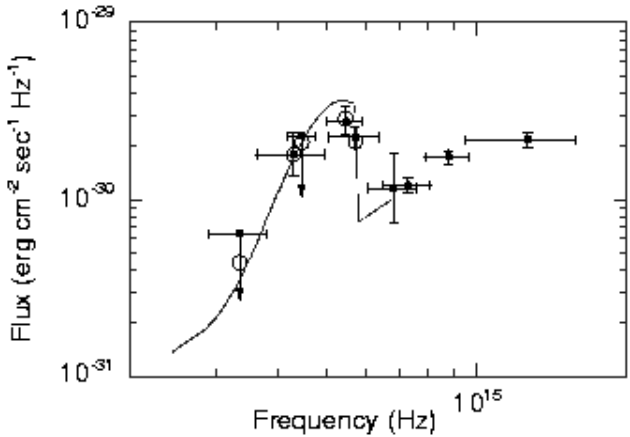,height=6cm,clip=}}}
\caption{Radiation flux ($erg~cm^{2} s^{-1} Hz^{-1}$) versus frequency (Hz) in 
the
region near the cyclotron resonance frequency. The data of Fig.1 (close
symbols) are compared 
with the  the ion cyclotron
emission computed on the basis of our model  (solid line).
In order to make the  curve immediately comparable to the data points,
we have filtered the curve
through the passbands actually used in the photometric observations.
The resulting fluxes are shown as open symbols.}
\end{figure}

Blackbody  emission conditions could be met  in a lower density region
of the outer  part of the  atmosphere,  with a slightly larger  plasma
thickness.  This,   however, would not   change the main  issue of the
present model: the observed  Geminga   spectral feature comes  from  a
plasma which belongs to   the star atmosphere.  One may  alternatively
think  that the cyclotron line was  emitted  by the low density matter
surrounding the  star. This would imply that  what is observed depends
upon a process  of photon collection  over  a wide plasma  volume and,
consequently, over a wide range  of cyclotron frequencies. 
However, this is  contradicted by the comparatively
well defined peak now observed.

\subsection{Emission Line Broadening Mechanism.}

The   real   part  of    $n^{2}(\omega,\theta)$ shown in Fig.2
exhibits a very  narrow frequency band of  anomalous dispersion.  This
indicates that  the width  of the  observed   emission line cannot  be
explained in terms of the thermal Doppler broadening ($\Delta \omega 
\simeq \omega | n cos \theta |  \frac{v_{thi}}{c}$, where $v_{thi}$ is
the ion thermal velocity). 
This is by  far insufficient to fit  the  observations in
the range  of temperatures required   to explain the intensity  of the
line. Another  possible source of  broadening are collisional
effects. These, however, have been shown to be insufficient to account
for the width  of the cyclotron feature. The   only mechanism left  to
account for  such an  observed width is   the structure of  a dipole-like
magnetic  field  associated with  the  star.  In  the case  of a perfect
magnetic dipole located  at the centre of the  star,  the intensity of
the magnetic field increases  by a factor  two when moving on the star
surface  from the (magnetic) equator to  the  poles. As a consequence,
the ion cyclotron frequency should also change by a factor two, if the
emitting plasma were to cover the whole star surface. The shape of the
emission line is  then  determined by  the superposition of  radiation
emitted by regions  of the star with  different magnetic field  values
and possibly different plasma temperature.  The  intensity of the line
in  our model is obtained  as the integral  of the  emission from each
point of the star surface: 

$$
I(\omega) = \frac{1}{D^2} \int_\Sigma i(\omega,\omega_{ci}) \frac{\vec
n \bullet \vec k}{|\vec k|} d^{2}\Sigma \eqno  (6)
$$

where $D$ is the distance of the star from the observer and the integral is
extended to the whole surface, $\Sigma$, taking into account the usual geometry
effects ($\vec n$ represents the unit vector perpendicular to
$d^{2}\Sigma$, while $\vec k$ is the propagation vector pointing toward the 
observer).
The non-homogeneous emission also depends on the relative position of the
observer with respect to the star rotation axis and, to properly account for
the shape and the intensity of the feature, on the angle $\gamma$ between the
rotation and magnetic axis. The computed spectrum, of course, will be
averaged over the star rotation period. These model assumptions also allow
us to determine the modulation depth expected for the observed spectral
feature, yielding a clear observational test.

\section{Model and Interpretation.}

In the context of the model described above, the intensity of the emission
line depends upon 1) temperature of the cyclotron-emitting plasma 2)
emitting fraction of the star surface. The range of the dipole field
spanned by the emitting plasma surface determines the frequency width of
the observed feature.
Clearly, the ion gyrofrequency formula allows us to determine the value of
the star' s magnetic field as a function of the frequency of the observed
line.
Since the ratio A/Z (which defines the chemical composition and the
ionization level of the emitting atmosphere) is unknown, the magnetic field
value is, in principle, determined only within a factor of $\sim$ 2. On the 
other
hand, our model yields a clear prediction of a sharp intensity decrease at
the frequency value corresponding to the B field maximum, located close to
the magnetic poles.
To compute a B field value, the emitting medium is assumed to be either H
or He, yielding a well-defined A/Z ratio. This is consistent with the
strong stratification of the elements induced by the huge gravitational
field of the neutron star. Hydrogen and Helium differ by a factor two in
their ratio A/Z, so that the cyclotron second harmonic of the heavier
element overlaps exactly the fundamental one of the lighter. 
The position of the polar caps with respect to the star magnetic axis is
shown in Fig.3. The polar cap extension, given by the  angle $\beta$, determines
the frequency width of the cyclotron emission. The magnetic axis forms an
angle $\gamma$ with the rotation axis. The model will be fully determined once 
the
angle, $\alpha$, between the observer and the rotation axis, is
defined (see Fig.4). Fig.5 compares the  data (filled circles) to 
the 
cyclotron emission model (open circles). The thin continous line shows the
detailed shape of the feature as determined by the model. The open circles
are obtained by integrating the model data over the filter passbands. The
computed magnetic field ranges from $3.8~ 10^{11}~ G$ for the case of a pure
Hydrogen plasma to $7.6~ 10^{11} G$ in the case of Helium.
The model data shown in Fig. 5 have been obtained with the following
assumptions: magnetic pole plasma temperature $T_{0} = 9~ 10^{7}~^{\circ}K$; 
temperature
profile along the polar
caps assumed to be gaussian-like $T = T_{0}
exp[-(\beta/\beta_{0})^{4}]$ with 
$\beta_{0} =57^{\circ}$.
 With
this choice of parameters the plasma temperature drops to 1/10 of its
maximum value in about $60^{\circ} $. Such an extension of the plasma polar 
caps is
required to explain the width of the line. Of course, the observed
line intensity and the plasma temperature are also determined by
Geminga's radius and by its distance from the observer. We assumed a radius 
$r_{0} =10~km$ and the parallax distance of  157 pc. Reasonable geometry uncertainties,
however, do not change the order of magnitude of the plasma temperature
required to emit such an intense cyclotron line.
Fig.6 gives a prediction of our model under the assumption of the
oblique rotation geometry ($\gamma = 90^{\circ}$) and for $\alpha$ close to 
$20^{\circ}$,
both the 
line
intensities and profiles are seen to vary with the rotation phase. For this
choice of parameters the modulation factor is about 15\% and the profile is
seen to sharpen close to the emission maximum (i.e. when the $\vec B$ axis sweeps
over the observer direction). Obviously, the region around $5500 \AA$ should be
the ideal one for observing the feature modulation and profile.

\begin{figure}
\centerline{\hbox{\psfig{figure=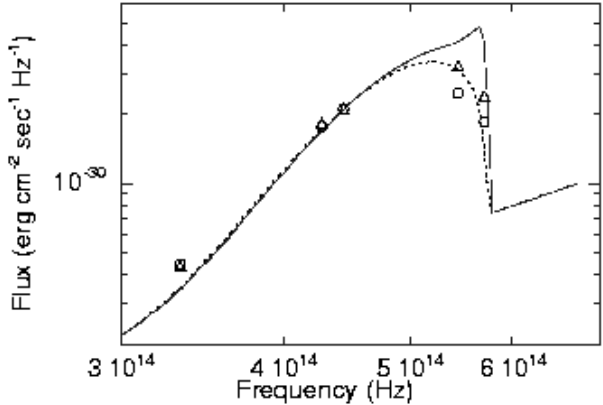,height=6cm,clip=}}}
\caption{Prediction of the model under the assumption of the oblique
rotation geometry ($\gamma=90$) and for $\alpha$ close to $20^{\circ}$. Both 
the line intensities
(triangles and circles refer to the maximum and minimum emission
respectively) and profiles (solid and dotted lines) are seen to vary with
the rotation phase.}
\end{figure}

\section{Electron Cyclotron Emission and Absorption.}

Geminga's polar cap atmosphere has been modelled as a fully ionized gas with a
density  $\simeq 10^{19} cm^{-3}$ and temperature  $\simeq 9~10^{7}~^{\circ}K$. 
The associated
electron-ion energy equipartion time, under these model assumptions,
keeps ions and
electrons close in temperature. The electron cyclotron emission process
is thus quite efficient and the electron cyclotron line, associated to the
companion ion cyclotron line, is expected to be an observable feature. This
is in contradiction with the experimental data (see the combined ROSAT/ASCA X-ray spectrum)
that show no line feature at E=4 or 8 keV  i.e. in the X-rays spectral
range where the electron spectral line should fall.
A possible explanation based on the observation that, within a
blackbody emission approximation, the power emitted at the electron
cyclotron frequency is so high that the rate of relaxation between
parallel and perpendicular (to the magnetic field) temperature is not
sufficient to keep the electron distribution function isotropic
(Ichimaru 1973; Trubnikov 1965). The anisotropy of the electron
distribution function, following this line of thought, would then
reach a steady state when the cyclotron emitted power, which is
proportional to the perpendicular temperature (Bornatici et al. 1983), is lower
than the one predicted by the isotropic case. 

In   order  to  give   a quantitative estimate   of  the  steady state
anisotropy and emission a kinetic calculation is required by means of a
relativistic Fokker-Planck equation which  includes a quasilinear term
of   interaction with  electron  cyclotron  radiation.   This  is,  at
present, beyond the target of this work. 

A possible
explanation of the fact that the residual emission is not observed  
is given by the polar cap heating model proposed
for Geminga by Halpern \& Ruderman (1993). Following this model  a flux of  
$e^{+} e^{-}$
pairs is created on closed field lines lying outside the star and
channelled into the polar caps of Geminga with a residual energy of about
6.5 erg each (Halpern \& Ruderman 1993). The $e^{\pm}$ cloud embedded in the
dipole magnetic field, which decreases by increasing the distance from the
star surface, can thus act as a "second harmonic" resonant absorber of
cyclotron radiation emitted from regions closer to the star surface. The
resonant radial position is located at $r_{2nd}=2^{1/3}r_{0}$, and, owing to the
electron cyclotron line width $\Delta \omega/\omega_{ce} = v_{the}/c$ (where 
$\omega_{ce}$ is the angular
frequency of the electron cyclotron emission and $v_{the}$ the thermal velocity
of the electrons) extends over about 300 m.
It can be shown (Bornatici et al. 1983) that the E=4 keV line is
efficiently absorbed ("optical depth" $\tau \simeq 16$) by the $e^{\pm}$ 
cloud, providing that ($n_{\pm}T_{\pm}$) $\simeq 100$  where $n_{\pm}$  and
$T_{\pm}$ are density and temperature of  $e^{\pm}$ respectively
($n_{\pm}$ in $10^{15}$ units and $T_{\pm}$ in keV).
Typical densities $n_{\pm} \simeq 10^{17} cm^{-3}$, corresponding to column 
densities $\simeq 10^{22}cm^{-2}$, with a temperature of $\simeq  2~ 10^{6}~^{\circ}K$ 
(0.2 keV), for instance, will
attenuate the electron cyclotron line by a factor $\simeq 10^{7}$. The electron
cyclotron emission, following these model assumptions, can no longer be
observed as a spectral feature.

\section{Conclusions.}

The data shown in Fig.1
leave no room for doubt that a wide emission feature exists
in the optical region of Geminga's thermal continuum. The feature falls in
the wavelength region where the atmospheric ion-cyclotron emission will be
located close to the surface of a magnetic neutron star. Since Geminga is a
magnetic neutron star, to wit its periodic $\gamma$-ray emission, and most
probably has an atmosphere, to wit its soft X-ray emission, we have
provided here a semi-quantitative interpretation for such feature. It is
based on the reasonable assumption that the polar cap regions of the NS are
covered by a thin plasma layer heated to a temperature higher that the
global surface atmosphere by, e.g., infalling particles. This is not a new
scenario per se. It was foreseen both in the case of INS accretion of
ionized matter funnelled towards the poles by the B-field
configuration 
and
of magnetospheric particles drawn back to the polar surface by the strong 
E field
induced by the oblique rotator. \\
The plausibility of this emission model in the visible range frequencies
is also supported by estimate of power balance performed
along the line proposed by Halpern \& Ruderman. In the case of Geminga a
pair flux in excess of $\dot N = 10^{38} s^{-1}$  can release in the
emitting 
plasma
layer a linear power density $\dot N dE/dr \simeq 10^{28} erg~cm^{-1}
s^{-1}$ (Jackson 1975).
This power is sufficient to compensate plasma losses mainly due to
the ion cyclotron emission  and Bremsstrahlung over the whole star surface.\\
What is new here is the excellent fit
obtained to the multiple experimental data by our physical model using a
minimum of assumption. In particular, we have shown that the feature could
not originate over the whole star surface, because global B-field
variations would induce a  feature wider than observed. The only free
parameter is the geometry of the emission with respect to the observer;
note, however, that our geometry is fully compatible with the oblique
rotator proposed for Geminga by Halpern \& Ruderman (1993).
The assumption that the composition of the outer emitting layer is either H
or a light, fully ionized element mixture is supported by the estimated
value for the magnetic field. Such a value is in good agreement with the
standard pulsar magnetic field prediction. It represents, in fact, the
first independent measurement of the surface magnetic field of an INS.

\begin{acknowledgements}
Useful discussions with Prof. Bruno Bertotti are gratefully acknowledged.
\end{acknowledgements}

\end{document}